\RequirePackage[mathlines]{lineno} 
\documentclass[prd,twocolumn,showpacs,amsmath,amssymb]{revtex4}
\usepackage{overpic,graphicx}
\usepackage{dcolumn}
\usepackage{bm}
\usepackage{rotating}
\usepackage{subfigure}
\usepackage{color}

\begin{document}

\title{\bf \boldmath
Improved measurement of the absolute branching fraction of $D^{+}\rightarrow \bar K^0 \mu^{+}\nu_{\mu}$}

\author{
  M.~Ablikim$^{1}$, M.~N.~Achasov$^{9,e}$, X.~C.~Ai$^{1}$,
  O.~Albayrak$^{5}$, M.~Albrecht$^{4}$, D.~J.~Ambrose$^{44}$,
  A.~Amoroso$^{49A,49C}$, F.~F.~An$^{1}$, Q.~An$^{46,a}$,
  J.~Z.~Bai$^{1}$
% ...
  \\
  \vspace{0.2cm}
  (BESIII Collaboration)\\
  \vspace{0.2cm} {\it
    $^{1}$ Institute of High Energy Physics, Beijing 100049, People's Republic of China\\
% ...
$^{4}$ Bochum Ruhr-University, D-44780 Bochum, Germany\\
    $^{5}$ Carnegie Mellon University, Pittsburgh, Pennsylvania 15213, USA\\
% ...
$^{9}$ G.I. Budker Institute of Nuclear Physics SB RAS (BINP), Novosibirsk 630090, Russia\\
% ...
$^{44}$ University of Rochester, Rochester, New York 14627, USA\\
% ...
$^{46}$ University of Science and Technology of China, Hefei 230026, People's Republic of China\\
% ...
$^{49}$ (A)University of Turin, I-10125, Turin, Italy; (B)University of Eastern Piedmont, I-15121, Alessandria, Italy; (C)INFN, I-10125, Turin, Italy\\
% ...
$^{a}$ Also at State Key Laboratory of Particle Detection and Electronics, Beijing 100049, Hefei 230026, People's Republic of China\\
% ...
$^{e}$ Also at the Novosibirsk State University, Novosibirsk, 630090, Russia\\
% ...
  }
}

\begin{abstract}
By analyzing 2.93 fb$^{-1}$ of data collected at $\sqrt s=3.773$ GeV
with the BESIII detector, we measure the
absolute branching fraction
${\mathcal B}(D^{+}\rightarrow\bar K^0\mu^{+}\nu_{\mu})=(8.72 \pm
0.07_{\rm stat.} \pm 0.18_{\rm sys.})\%$,
which is consistent with previous measurements within uncertainties but with significantly improved precision.
Combining the Particle Data Group values of ${\mathcal B}(D^0\to K^-\mu^+\nu_\mu)$, ${\mathcal B}(D^{+}\rightarrow\bar K^0 e^{+}\nu_{e})$,
and the lifetimes of the $D^0$ and $D^+$ mesons with the value of ${\mathcal B}(D^{+}\rightarrow\bar K^0 \mu^{+}\nu_{\mu})$ measured in this work,
we determine the following ratios of partial widths:
$\Gamma(D^0\to K^-\mu^+\nu_\mu)/\Gamma(D^{+}\rightarrow\bar
K^0\mu^{+}\nu_{\mu})=0.963\pm0.044$ and $\Gamma(D^{+}\rightarrow\bar
K^0 \mu^{+}\nu_{\mu})/\Gamma(D^{+}\rightarrow\bar K^0
e^{+}\nu_{e})=0.988\pm0.033$.

\end{abstract}


\maketitle

\oddsidemargin  -0.2cm
\evensidemargin -0.2cm

\section{Introduction}

Experimental studies of $D$ semileptonic decays provide helpful information to understand
$D$ decay mechanisms. Their decay branching fractions (${\mathcal B}$)
can serve to test isospin conservation and leptonic universality in $D$ semileptonic decays.
Isospin conservation implies that the partial widths ($\Gamma$)
of $D^{0}\rightarrow K^{-}\mu^{+}\nu_{\mu}$ and $D^{+}\rightarrow\bar K^0\mu^{+}\nu_{\mu}$
should be equal.
Furthermore, Ref.~\cite{zphyc} predicts that
$\Gamma(D\rightarrow\bar{K}\mu^{+}\nu_{\mu})$ is less than
$\Gamma(D\rightarrow\bar{K}e^{+}\nu_{e})$ by about 3\%
due to different form factors and phase space.
Using the branching fractions and the lifetimes of the $D^0$ and $D^+$ mesons ($\tau_{D^0}$, $\tau_{D^+}$),
taken from the Particle Data Group (PDG) \cite{pdg2014},
we obtain
$\Gamma(D^0\to K^-\mu^+\nu_\mu)/\Gamma(D^{+}\rightarrow\bar K^0\mu^{+}\nu_{\mu})=0.91\pm0.07$
and
$\Gamma(D^{+}\rightarrow\bar K^0\mu^{+}\nu_{\mu})/\Gamma(D^{+}\rightarrow\bar K^0e^{+}\nu_{e})=1.04\pm0.07,$
where the uncertainties are dominated by ${\mathcal B}(D^{+}\rightarrow\bar K^0\mu^{+}\nu_{\mu})$ \cite{pdg2014}.
Thus, an improved measurement of ${\mathcal B}(D^{+}\rightarrow\bar K^0\mu^{+}\nu_{\mu})$ will
be helpful to understand $D$ decay mechanisms with better accuracy.
In addition, the improved ${\mathcal B}(D^+\to \bar K^0 \mu^{+}\nu_\mu)$
can also be used to precisely determine the form factor $f^K_+(0)$ and the
quark mixing matrix element $|V_{cs}|$ from $D$ semileptonic decays~\cite{rongg}.

Previous measurements of ${\mathcal B}(D^{+}\rightarrow\bar K^0\mu^{+}\nu_{\mu})$ come from MARKIII~\cite{mark3}, FOCUS~\cite{FOCUS} and BESII~\cite{BESII}.
In this paper, by analyzing 2.93 fb$^{-1}$ of data ~\cite{BESIII292} collected
at the center-of-mass energy of $\sqrt s=3.773$ GeV by the BESIII detector ~\cite{BESIII},
we determine the absolute branching fraction of
$D^{+}\rightarrow\bar K^0\mu^{+}\nu_{\mu}$.
Throughout the paper, charge conjugation is implied.

\section{BESIII detector and Monte Carlo}

The BESIII
detector is a cylindrical detector with a solid-angle coverage of 93\% of $4\pi$
that operates at the BEPCII collider. It consists of several main components.
A 43-layer main drift chamber (MDC) surrounding the beam pipe performs precise determinations
of charged particle trajectories and provides a measurement of the specific ionization energy loss ($dE/dx$)
that is used for charged particle identification (PID). An array of time-of-flight counters
(TOF) is located radially outside the MDC and provides additional PID information.
A CsI(Tl) electromagnetic calorimeter (EMC)
surrounds the TOF and is used to measure the energies of photons and
electrons. A solenoidal superconducting magnet located outside the EMC
provides a 1 T magnetic field in the central tracking region of the
detector. The iron flux return of the magnet is instrumented with about
1272 m$^2$ of resistive plate muon counters (MUC) arranged in nine
layers in the barrel and eight layers in the endcaps that are used
to identify muons with momentum greater than 0.5 GeV/$c$.
More details about the BESIII detector are described in Ref.~\cite{BESIII}.

A GEANT4-based~\cite{geant4} Monte Carlo (MC) simulation software package, which includes the geometric
description of the detector and its response, is used to determine
the detection efficiency and to estimate the potential backgrounds.
An inclusive MC sample, which includes
the $D^0\bar D^0$, $D^+D^-$, and non-$D\bar D$ decays of $\psi(3770)$,
the initial state radiation (ISR) production of $\psi(3686)$ and $J/\psi$,
the $q\bar q$ ($q=u$, $d$, $s$) continuum process,
the Bhabha scattering events, and
the di-muon and di-tau events,
is produced at $\sqrt s=3.773$ GeV.
The $\psi(3770)$ decays are generated by the MC generator KKMC \cite{kkmc},
in which ISR effects \cite{isr} and
final state radiation (FSR) effects \cite{photons} are simulated.
The known decay modes of the charmonium states
are generated using EvtGen \cite{evtgen} with the branching fractions
set to PDG values \cite{pdg2010},
and others are generated using LundCharm \cite{lundcharm}.
The $D^+ \to \bar K^0 \mu^+\nu_\mu$ signal is simulated with the
modified pole model~\cite{BK-model}.

\section{Method}
In $e^{+}e^{-}$ collisions at $\sqrt{s}=3.773$ GeV, the $\psi$(3770)
resonance decays predominately into a $D^0\bar{D}^0$ or a $D^+D^-$
pair. In an event where a $D^{-}$ meson (called the single tag (ST)
$D^{-}$ meson) is fully reconstructed, the presence of a $D^{+}$ meson is
guaranteed. In the systems recoiling against the ST $D^-$ mesons, we can
select the semileptonic decays of $D^{+}\rightarrow\bar
K^0\mu^{+}\nu_{\mu}$ (called the double tag (DT) events).
For a special ST mode $i$, the ST and DT yields observed in data are given by
\begin{equation}
N^{i}_{\rm ST} = 2 N_{D^+D^-} {\mathcal B}^i_{\rm ST} \epsilon^i_{\rm ST},
\end{equation}
and
\begin{equation}
N^{i}_{\rm DT} = 2 N_{D^+D^-} {\mathcal B}^i_{\rm ST} {\mathcal B}(D^{+}\rightarrow\bar K^0\mu^{+}\nu_{\mu})
\epsilon^i_{{\rm ST},D^{+}\rightarrow\bar K^0\mu^{+}\nu_{\mu}},
\end{equation}
where $N_{D^+D^-}$ is the number of $D^+D^-$ pairs produced in data,
${\mathcal B}^i_{\rm ST}$ and ${\mathcal B}(D^{+}\rightarrow\bar K^0\mu^{+}\nu_{\mu})$
are the branching fractions for
the ST mode $i$ and the $D^{+}\rightarrow\bar K^0\mu^{+}\nu_{\mu}$ decay,
$\epsilon^i_{\rm ST}$ is the efficiency of reconstructing the ST mode $i$ (called the ST efficiency),
and $\epsilon^i_{{\rm ST},D^{+}\rightarrow\bar K^0\mu^{+}\nu_{\mu}}$ is the efficiency of simultaneously
finding the ST mode $i$ and the $D^{+}\rightarrow\bar K^0\mu^{+}\nu_{\mu}$ decay (called the DT efficiency).
Based on these two equations, the absolute branching fraction for
$D^{+}\rightarrow\bar K^0\mu^{+}\nu_{\mu}$
can be determined by
\begin{equation}
{\mathcal B}(D^{+}\rightarrow\bar K^0\mu^{+}\nu_{\mu}) = \frac{N^{\rm tot}_{\rm DT}}{N^{\rm tot}_{\rm ST}\bar \epsilon_{D^{+}\rightarrow\bar K^0\mu^{+}\nu_{\mu}}},
\end{equation}
where
$\bar \epsilon_{D^{+}\rightarrow\bar K^0\mu^{+}\nu_{\mu}} =
\sum_i (N^i_{\rm ST} \epsilon^i_{{\rm ST},D^{+}\rightarrow\bar K^0\mu^{+}\nu_{\mu}}/\epsilon^i_{\rm ST})/N^{\rm tot}_{\rm ST}$
is the averaged efficiency of reconstructing the $D^{+}\rightarrow\bar K^0\mu^{+}\nu_{\mu}$ decay by the ST yields in data.

\section{ST $D^{-}$ mesons}
\label{sec:evtsel}

The ST $D^{-}$ mesons are reconstructed using six hadronic decay modes:
$K^{+}\pi^{-}\pi^{-}$, $K^0_{S}\pi^{-}$, $K^{+}\pi^{-}\pi^{-}\pi^{0}$,
$K^0_{S}\pi^{-}\pi^{0}$, $K^0_{S}\pi^{+}\pi^{-}\pi^{-}$
and $K^{+}K^{-}\pi^{-}$. The decays of $K^0_{S}$ and $\pi^0$ mesons are
identified in $K^0_{S}\to \pi^{+}\pi^{-}$ and $\pi^0\to\gamma\gamma$, respectively.

All charged tracks used in this analysis are required to be within
a polar-angle ($\theta$) range of $|\rm{cos~\theta}|<0.93$.
Except for those from $K^0_{S}$ decays, all tracks are required to originate
from an interaction region defined by
$V_{xy}<$ 1.0 cm and $|V_{z}|<$ 10.0 cm,
where $V_{xy}$ and $|V_{z}|$ refer to the distances of closest approach
of the reconstructed track to the Interaction Point (IP) in
the $xy$ plane and the $z$ direction (along the beam), respectively.

The charged kaons and pions are
identified by the $dE/dx$ and TOF information.
The combined Confidence Levels
for pion and kaon hypotheses ($CL_{\pi}$ and $CL_{K}$) are calculated, respectively.
A charged track is identified as a kaon (pion) if the confidence levels
satisfy $CL_{K}>CL_{\pi}$ ($CL_{\pi}>CL_{K}$).

The charged tracks from $K^0_{S}$ decays are required to
satisfy $|V_{z}|<$ 20.0 cm.
The two oppositely charged tracks are assigned
as $\pi^+\pi^-$ without PID. The $\pi^+\pi^-$ pair is constrained to
originate from a common vertex and is required to have an invariant mass
within $|M_{\pi^{+}\pi^{-}} - M_{K_{S}^{0}}|<$ 12 MeV$/c^{2}$, where
$M_{K_{S}^{0}}$ is the $K^0_{S}$ nominal mass \cite{pdg2014}. The
$K^0_S$ candidate is required to have a decay length larger than 2
standard deviations of the vertex resolution away from the IP.

Photon candidates are selected using the information from the EMC.
It is required that the shower time be within 700 ns of the event start time,
the shower energy be greater than 25 (50) MeV
if the crystal with the maximum deposited energy in that cluster
is in the barrel (endcap) region~\cite{BESIII},
and the opening angle between the candidate shower and
any charged tracks be greater than $10^{\circ}$.
To reconstruct $\pi^0$, the invariant mass of the accepted $\gamma\gamma$
pair is required to be within $(0.115, 0.150)$ GeV$/c^{2}$.
To improve resolution, a kinematic fit is performed to constrain
the $\gamma\gamma$ invariant mass to the $\pi^{0}$ nominal mass \cite{pdg2014}.

To identify the ST $D^{-}$ mesons, we define two variables,
the energy difference $\Delta E=E_{mKn\pi} - E_{\rm beam}$ and
the beam energy constrained mass $M_{\rm BC} = \sqrt{E^{2}_{\rm beam}-|\vec{p}_{mKn\pi}|^{2}}$
of the $mKn\pi$ ($m=1, 2$; $n=1, 2, 3$) final states,
where $E_{\rm beam}$ is the beam energy,
$\vec{p}_{mKn\pi}$ and $E_{mKn\pi}$ are the measured momentum and energy of
the $mKn\pi$ final state in the $e^+e^-$ center-of-mass frame.
For each ST mode, if there is more than one combination surviving,
only the one with the minimum $|\Delta E|$ is kept. To suppress combinatorial backgrounds,
$\Delta E$ is required to be
within $(-25, +25)$ MeV for the $K^{+}\pi^{-}\pi^{-}$, $K^0_{S}\pi^{-}$,
$K^0_{S}\pi^{+}\pi^{-}\pi^{-}$ and $K^{+}K^{-}\pi^{-}$ final states, and
be within $(-55, +40)$ MeV for the $K^{+}\pi^{-}\pi^{-}\pi^{0}$ and $K^0_{S}\pi^{-}\pi^{0}$ final states.

\begin{figure}[htp]
  \centering
  \includegraphics[width=2.8in]{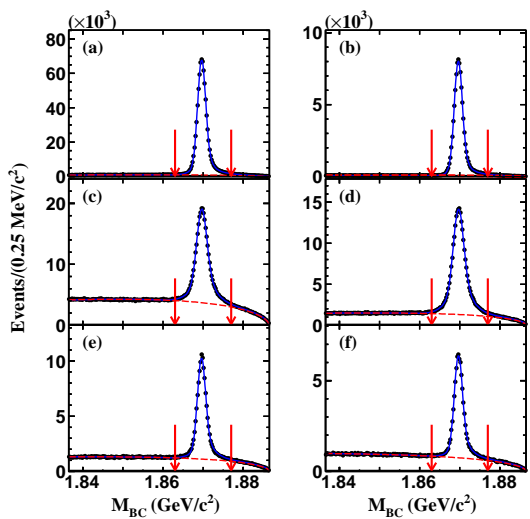}
  \caption{
  (color online) Fits to the $M_{\rm BC}$ distributions of
  (a) $K^{+}\pi^{-}\pi^{-}$,
  (b) $K^0_{S}\pi^{-}$,
  (c) $K^{+}\pi^{-}\pi^{-}\pi^{0}$,
  (d) $K^0_{S}\pi^{-}\pi^{0}$,
  (e) $K^0_{S}\pi^{+}\pi^{-}\pi^{-}$ and
  (f) $K^{+}K^{-}\pi^{-}$ combinations.
The dots with error bars are data,
the blue solid curves are the fit results,
the red dashed curves are the fitted backgrounds and
the pair of red arrows in each sub-figure denote the ST $D^-$ signal region.
  }\label{fig:datafit_Massbc}
\end{figure}

To obtain the ST yield, we apply a fit to the $M_{\rm BC}$ distributions
of the accepted $mKn\pi$ final states for data.
In the fits,
the $D^{-}$ signal is modeled by a MC-determined shape of the $M_{\rm BC}$ distribution
convoluted with a double Gaussian function
and the combinatorial background shape is described by the
ARGUS function \cite{ARGUS}.
The fit results are shown in Figure \ref{fig:datafit_Massbc}.
The candidates with $M_{\rm BC}$ in the range $(1.863,1.877)$ GeV/$c^2$
(signal region) are kept for further analysis.
The ST yields and the ST efficiencies estimated from the inclusive MC sample
are summarized in Table \ref{tab:singletagN}.
The total ST yield is $N^{\rm tot}_{\rm ST}= 1522474 \pm 2215$.

\begin{table*}[htp]
  \centering
  \caption{Summary of
   the ST yields ($N^i_{\rm ST}$),
   the ST and DT efficiencies ($\epsilon^i_{\rm ST}$ and $\epsilon^i_{\rm DT}$),
   and the efficiencies of detecting $D^{+}\rightarrow\bar K^0\mu^{+}\nu_{\mu}$
   ($\epsilon^i_{D^{+}\rightarrow\bar K^0\mu^{+}\nu_{\mu}}$).
   The efficiencies (in percent) do not include ${\mathcal B}(\pi^0\to\gamma\gamma)$ and
   ${\mathcal B}(\bar K^0\to \pi\pi)$.
   ${+-}$ and ${00}$ denote the $D^{+}\rightarrow\bar K^0\mu^{+}\nu_{\mu}$ signals, which are
   reconstructed via $\bar K^0\to \pi^+\pi^-$ and $\bar K^0\to \pi^0\pi^0$, respectively.
   The DT efficiencies have been corrected according to the differences
   of the efficiencies of the $\mu^+$ tracking, the $\mu^+$ PID, the $\pi^0$ reconstruction of the signal
   side and the $E^{{\rm extra}~\gamma}_{\rm max}$ (see text) requirement between data and MC.
   The $i$ represents the $i$th ST mode.
   The uncertainties are statistical only.
   }\label{tab:singletagN}
\begin{tabular}{lcccccc}
  \hline
  Tag mode & $N^i_{\rm ST}$ & $\epsilon^i_{\rm ST}$ &
  $\epsilon^{i,+-}_{\rm DT}$ &
  $\epsilon^{i,+-}_{D^{+}\rightarrow\bar K^0 \mu^{+}\nu_{\mu}}$ &
  $\epsilon^{i,00}_{\rm DT}$ &
  $\epsilon^{i,00}_{D^{+}\rightarrow\bar K^0 \mu^{+}\nu_{\mu}}$ \\
  \hline
  $D^-\to K^{+}\pi^{-}\pi^{-}$            & 782669$\pm$\hspace{0.15cm}990  & 50.61$\pm$0.06 & 17.96$\pm$0.05 &35.49$\pm$0.11  & 10.75$\pm$0.06& 21.23$\pm$0.13\\
  $D^-\to K^0_{S}\pi^{-}$               & \hspace{0.15cm}91345$\pm$\hspace{0.15cm}320   & 50.41$\pm$0.17 & 18.66$\pm$0.16 &37.00$\pm$0.34  & 11.73$\pm$0.20& 23.26$\pm$0.40\\
  $D^-\to K^{+}\pi^{-}\pi^{-}\pi^{0}$     & 251008$\pm$1135 & 26.74$\pm$0.09 &\hspace{0.15cm}9.50$\pm$0.05 &35.52$\pm$0.23  & \hspace{0.15cm}5.17$\pm$0.06& 19.34$\pm$0.22\\
  $D^-\to K^0_{S}\pi^{-}\pi^{0}$        & 215364$\pm$1238 & 27.29$\pm$0.07 & 10.71$\pm$0.06 &39.23$\pm$0.24  &\hspace{0.15cm}6.11$\pm$0.07& 22.35$\pm$0.26\\
  $D^-\to K^0_{S}\pi^{+}\pi^{-}\pi^{-}$ & 113054$\pm$\hspace{0.15cm}889  & 28.31$\pm$0.12 &\hspace{0.15cm}9.98$\pm$0.08 &35.26$\pm$0.32  &\hspace{0.15cm}5.97$\pm$0.09& 21.08$\pm$0.34\\
  $D^-\to K^{+}K^{-}\pi^{-}$              & \hspace{0.15cm}69034$\pm$\hspace{0.15cm}460   & 40.83$\pm$0.24 & 13.34$\pm$0.14 &32.69$\pm$0.40  &\hspace{0.15cm}7.88$\pm$0.17& 19.31$\pm$0.43\\
  \hline
\end{tabular}
\end{table*}

\section{DT events}

From the surviving charged tracks and photons in the systems against the ST $D^-$ mesons,
the $D^+\to \bar K^0\mu^+\nu_\mu$ candidates are selected with
the following, optimized criteria.
The $\bar K^0$ is reconstructed using the decays
$\bar K^0\rightarrow\pi^{+}\pi^{-}$ and $\bar K^0\rightarrow\pi^{0}\pi^{0}$.
To select $\bar K^0(\pi^{0}\pi^{0})$,
the $\pi^{0}\pi^{0}$ invariant mass ($M_{\pi^0\pi^0}$) is required to be within
($0.45, 0.51$) GeV/$c^{2}$.
If more than one combination survives,
the one with the minimum
$\chi^2_1(\pi^0\to\gamma\gamma)+\chi^2_2(\pi^0\to\gamma\gamma)$
is kept,
where $\chi^2_1$ and $\chi^2_2$ are the chi-squares of the mass-constrained fits on
$\pi^0\to \gamma\gamma$.
The good charged tracks, photons, $\pi^0$, and $\bar K^0(\pi^+\pi^-)$ candidates are selected
with the same criteria as those used in the ST selection.

We require that there be only one good additional charged track
with charge opposite to that of the ST $D^-$ meson.
For muon identification, we combine the $dE/dx$, TOF and EMC information
to calculate the Confidence Levels
for electron, pion, kaon and muon hypotheses ($CL_{e}$, $CL_{\pi}$, $CL_{K}$, $CL_{\mu}$), respectively.
The charged track is assigned as a muon candidate if
the confidence levels satisfy
$CL_{\mu}>CL_{K}$, $CL_{\mu}>CL_{e}$ and $CL_{\mu}>$ 0.001.
To decrease the rate of mis-identifying pions as muons,
we require that
the energies deposited in the EMC by muons be within $(0.1, 0.3)$ GeV.

Since the neutrino is undetectable, we define a kinematic quantity
$$U_{\rm miss} \equiv  E_{\rm miss} - |\vec{p}_{\rm miss}|,$$
where $E_{\rm miss}$ and $|\vec{p}_{\rm miss}|$ are
the energy and momentum of the missing particle in the DT event, respectively.
$E_{\rm miss}$ is calculated by
$$E_{\rm miss} = E_{\rm beam} - E_{\bar K^0} - E_{\mu^+},$$
where
$E_{\bar K^0}$ and $E_{\mu^+}$ are the measured energies of $\bar K^0$
and $\mu^{+}$, respectively.
$\vec{p}_{\rm miss}$ is defined as
$$\vec{p}_{\rm miss} = |\vec{p}_{D^{+}} - \vec{p}_{\bar K^0} - \vec{p}_{\mu^+}|,$$
where $\vec{p}_{\bar K^0}$ and $\vec{p}_{\mu^+}$
are the measured momenta of $\bar K^0$ and $\mu^+$,
$\vec{p}_{D^{+}}$ is the constrained momentum of $D^+$ meson
$$\vec{p}_{D^{+}} =
(-\hat{p}_{D_{\rm ST}^{-}})
\sqrt{E_{\rm beam}^{2}-m_{D^{+}}^{2}},$$
where $\hat{p}_{D_{\rm ST}^{-}}$ is the momentum direction
of the ST $D^{-}$ meson and $m_{D^{+}}$ is the $D^{+}$ nominal mass \cite{pdg2014}.

\begin{figure}[htp]
  \centering
  \includegraphics[width=2.8in]{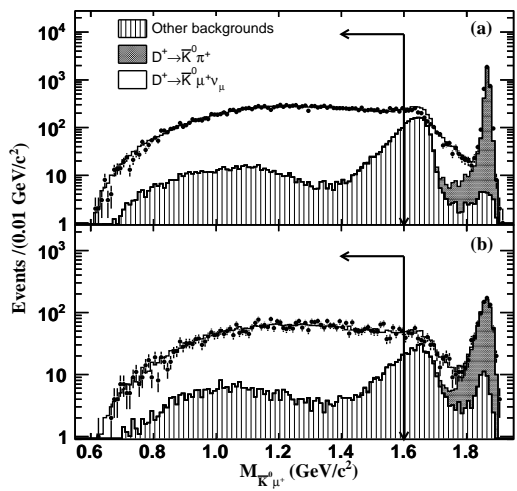}
  \caption{
  The $M_{\bar K^0\mu^+}$ distributions of the
(a) $D^{+}\rightarrow\bar K^0(\pi^+\pi^-)\mu^{+}\nu_{\mu}$ and
(b) $D^{+}\rightarrow\bar K^0(\pi^0\pi^0)\mu^{+}\nu_{\mu}$ candidates
of data (points with error bars) and the inclusive MC sample (histograms).
The perpendicular black arrow shows the requirement $M_{\bar K^0\mu^+}<1.6$ GeV/$c^2$.
Other backgrounds are dominated by $D^+ \to \bar K^0\pi^+\pi^0$.
For this figure,
$U_{\rm miss}$ is required to be within $(-0.06, +0.06)$ GeV.
}\label{fig:mk0mu}
\end{figure}

\begin{figure}[htp]
  \centering
  \includegraphics[width=2.8in]{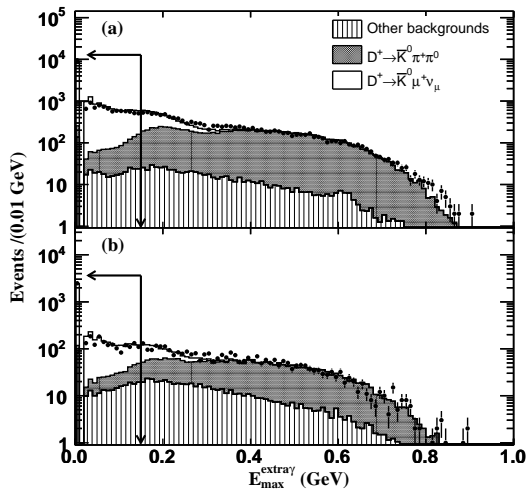}
  \caption{
  The $E^{{\rm extra}~\gamma}_{\rm max}$ distributions of the
(a) $D^{+}\rightarrow\bar K^0(\pi^+\pi^-)\mu^{+}\nu_{\mu}$ and
(b) $D^{+}\rightarrow\bar K^0(\pi^0\pi^0)\mu^{+}\nu_{\mu}$ candidates
of data (points with error bars) and the inclusive MC sample (histograms).
The perpendicular black arrow shows the requirement $E^{{\rm extra}~\gamma}_{\rm max}<0.15$ GeV.
For this figure, $U_{\rm miss}$ is required to be within $(-0.06, +0.06)$ GeV.
}\label{fig:egmax}
\end{figure}

Figures~\ref {fig:mk0mu} and \ref {fig:egmax} show the distributions of the $\bar K^0\mu^+$
invariant masses ($M_{\bar K^0\mu^+}$) and the maximum energies ($E^{{\rm extra}~\gamma}_{\rm max}$)
of any of the extra photons which have not been used in the DT event selection
from data and the inclusive MC sample,
respectively,
in which the backgrounds are dominated by $D^+ \to \bar K^0 \pi^+(\pi^0)$.
To suppress these backgrounds,
we require that the $D^+\to \bar K^0\mu^+\nu_\mu$ candidates
have $M_{\bar K^0\mu^+}<1.6$ GeV/$c^2$ and $E^{{\rm extra}~\gamma}_{\rm max}<0.15$ GeV.

The DT efficiency is determined by analyzing signal MC events.
Dividing $\epsilon_{\rm DT}$ by $\epsilon_{\rm ST}$, we obtain the efficiency of detecting
$D^{+}\rightarrow\bar K^0\mu^{+}\nu_{\mu}$ ($\epsilon^{+-}_{D^{+}\rightarrow\bar K^0 \mu^{+}\nu_{\mu}}$)
for each ST mode.
They are summarized in Table \ref{tab:singletagN}.
The averaged efficiencies of detecting
$D^{+}\rightarrow\bar K^0\mu^{+}\nu_{\mu}$ are determined to be
$$\bar \epsilon^{+-}_{D^{+}\rightarrow\bar K^0 \mu^{+}\nu_{\mu}}
= \frac{\sum_i (N^{i}_{\rm ST} \epsilon^{i,+-}_{D^{+}\rightarrow\bar K^0 \mu^{+}\nu_{\mu}} )}{N^{\rm tot}_{\rm ST}}
= (35.97\pm0.11)\%$$
and
$$\bar \epsilon^{00}_{D^{+}\rightarrow\bar K^0 \mu^{+}\nu_{\mu}}
= \frac{\sum_i (N^{i}_{\rm ST} \epsilon^{i,00}_{D^{+}\rightarrow\bar K^0 \mu^{+}\nu_{\mu}})}
       {N^{\rm tot}_{\rm ST}}
=(21.10\pm0.10)\%,$$
where
the $i$ denotes the sum over the six ST modes
and ${+-}$ and ${00}$ denote the $D^{+}\rightarrow\bar K^0\mu^{+}\nu_{\mu}$ signals,
which are
reconstructed via $\bar K^0\to \pi^+\pi^-$ and $\bar K^0\to \pi^0\pi^0$, respectively.

\begin{figure}[htp]
  \centering
  \includegraphics[width=2.8in]{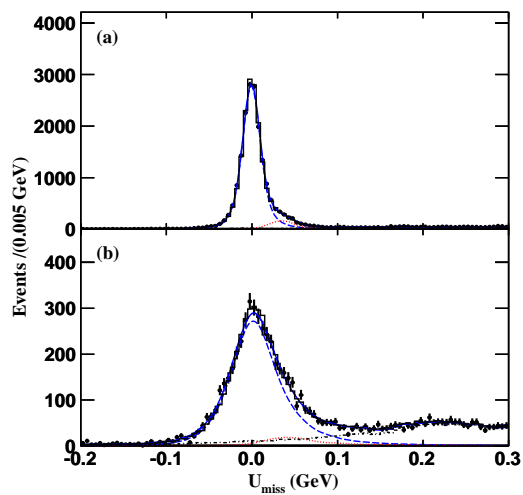}
  \caption{(color online) Fits to the $U_{\rm miss}$ distributions of the
(a) $D^{+}\rightarrow\bar K^0(\pi^+\pi^-)\mu^{+}\nu_{\mu}$ and
(b) $D^{+}\rightarrow\bar K^0(\pi^0\pi^0)\mu^{+}\nu_{\mu}$ candidates,
where
the histograms are the inclusive MC sample,
the dots with error bars are data,
the blue solid curves are the fit results,
the blue dashed curves are the $D^{+}\rightarrow\bar K^0\mu^{+}\nu_{\mu}$ signals,
the red dotted curves are the
$D^{+}\rightarrow\bar K^0\pi^{+}\pi^0$ peaking backgrounds and
the black dot-dashed curves are from other backgrounds.}
\label{fig:fit_Umistry1}
\end{figure}

To determine the signal yield, we perform simultaneous fits to the two $U_{\rm miss}$
distributions of the DT candidates, in which
$D^{+}\rightarrow\bar K^0\mu^{+}\nu_{\mu}$ is reconstructed via $\bar K^0\to \pi^+\pi^-$
and $\bar K^0\to \pi^0\pi^0$.
In the fits, we constrain the numbers of the
efficiency and branching fraction corrected DT events and
$D^+\to\bar K^0\pi^+\pi^0$ peaking backgrounds, respectively,
under the assumption that $K^0_S$ contributes to half of the neutral kaon decays.
We use MC-determined shapes convoluted with Gaussian functions
to describe the $D^{+}\rightarrow\bar K^0\mu^{+}\nu_{\mu}$ signal
and the $D^{+}\rightarrow\bar K^0\pi^{+}\pi^{0}$ peaking background,
and MC-based shape is also employed to represent the rest of the
background and their overall normalizations are free parameters in the fits.
The fit results are shown in Figure \ref{fig:fit_Umistry1}.
From the constrained fits, we determine the
efficiency and branching fraction corrected DT production yield in data to be
$$N^{\rm prd}_{\rm DT} = 132712 \pm 1041,$$
corresponding to the observed DT yields
$N^{+-,\rm obs}_{\rm DT} = 16516 \pm 130$ and $N^{00,\rm obs}_{\rm DT} = 4198 \pm 33$
for the $+-$ and $00$ modes, respectively.

\begin{figure}[htp]
  \centering
  \includegraphics[width=2.8in]{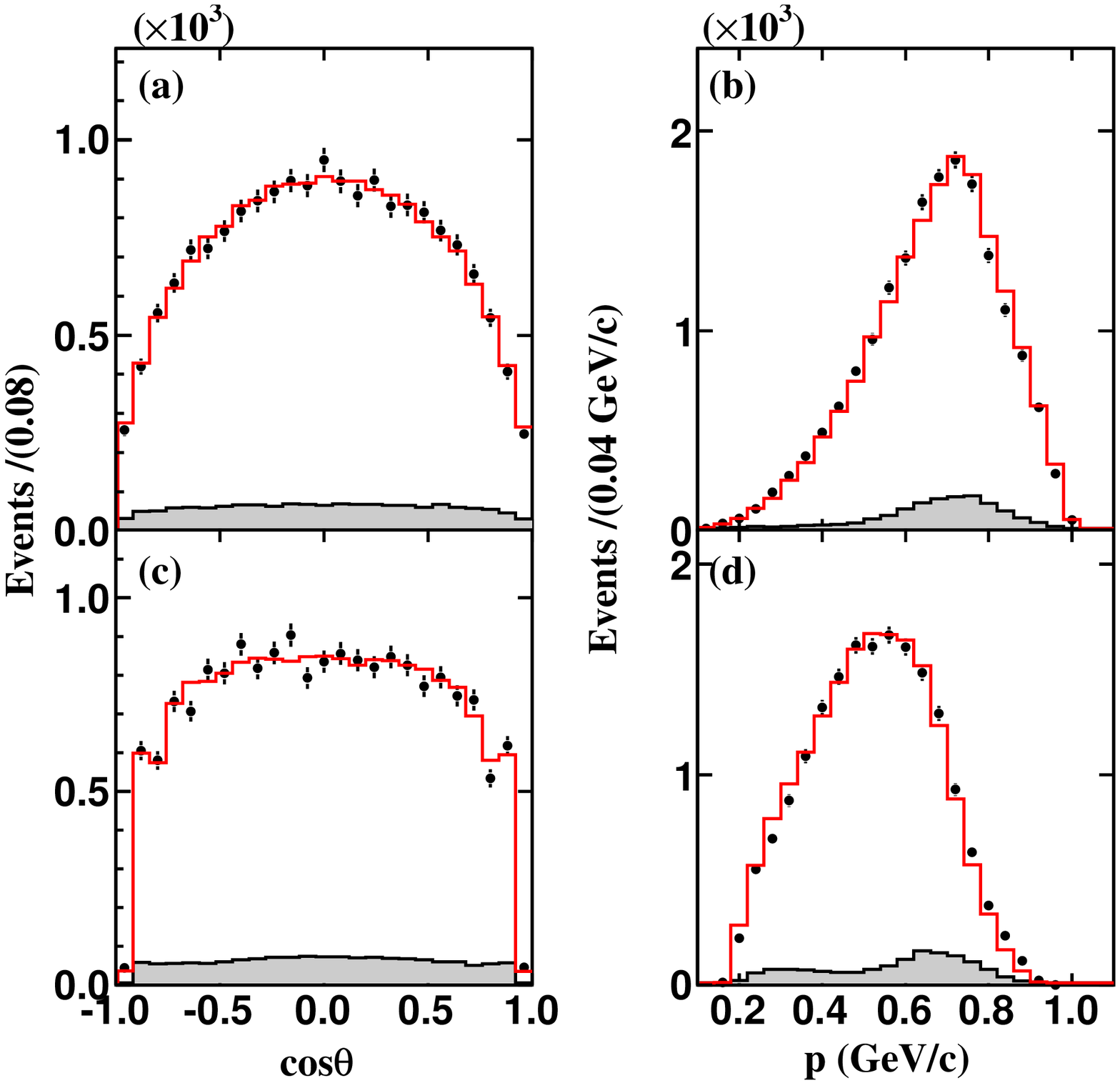}
  \caption{(color online)
  Comparisons of the $\cos\theta$ and momentum distributions of
  (a), (b) $\bar K^0$ and (c), (d) $\mu^+$ from the
  $D^{+}\rightarrow\bar K^0(\pi^+\pi^-)\mu^{+}\nu_{\mu}$ candidates,
  where the dots with error bars are data,
  the red histograms are the inclusive MC sample, and the hatched histograms are the MC
  simulated backgrounds.}\label{fig:comparisonpipm}
\end{figure}

\begin{figure}[htp]
  \centering
  \includegraphics[width=2.8in]{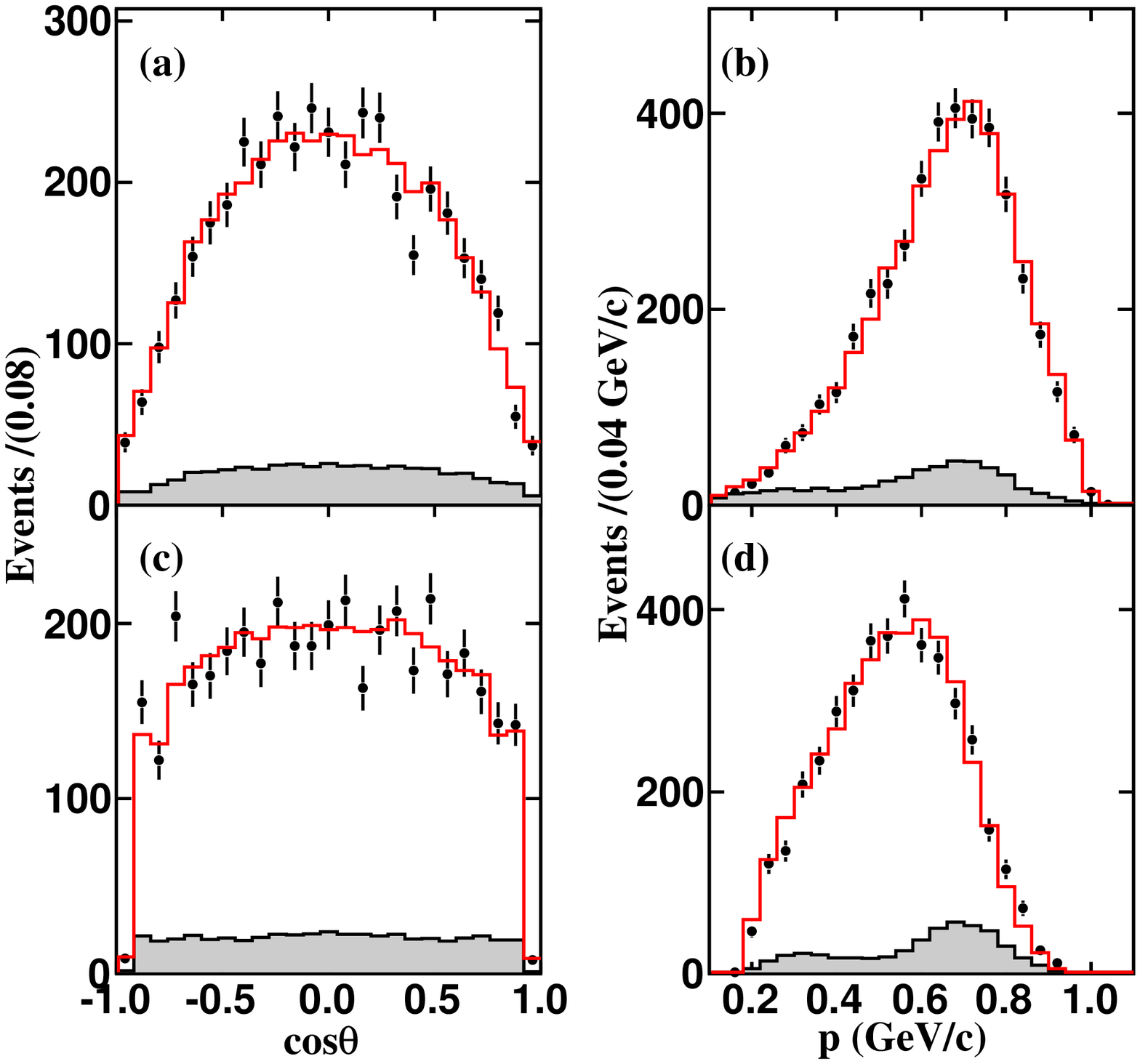}
  \caption{(color online)
  Comparisons of the $\cos\theta$ and momentum distributions of
  (a), (b) $\bar K^0$ and (c), (d) $\mu^+$ from the
  $D^{+}\rightarrow\bar K^0(\pi^0\pi^0)\mu^{+}\nu_{\mu}$ candidates,
  where the dots with error bars are data,
  the red histograms are the inclusive MC sample, and the hatched histograms are the MC
  simulated backgrounds. }\label{fig:comparison}
\end{figure}

\begin{figure}[htp]
\centering
\includegraphics[width=2.8in]{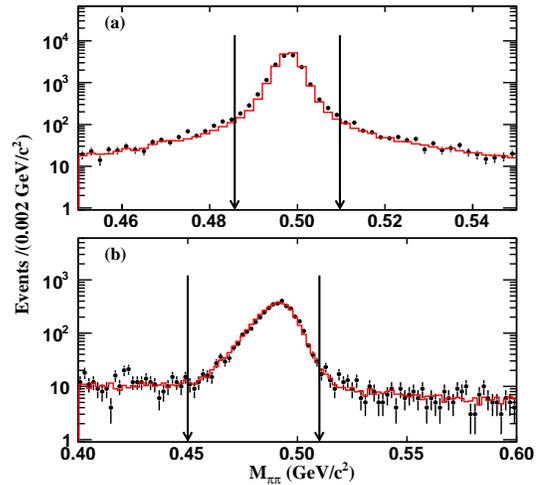}
\caption{(color online)
Comparisons of the (a) $M_{\pi^+\pi^-}$
and (b) $M_{\pi^0\pi^0}$ distributions of the
$D^{+}\rightarrow\bar K^0\mu^{+}\nu_{\mu}$ candidates,
where the dots with error bars are data,
the red histograms are the inclusive MC sample and
the arrow pairs denote the $K^0$ mass windows.}
\label{fig:mks}
\end{figure}

We compare the $\cos\theta$ and momentum distributions of
$\bar K^0$ and $\mu^+$ as well as the
$\pi\pi$ invariant mass spectra from the
$D^{+}\rightarrow\bar K^0(\pi^{+}\pi^{-})\mu^{+}\nu_{\mu}$ and
$D^{+}\rightarrow\bar K^0(\pi^{0}\pi^{0})\mu^{+}\nu_{\mu}$ candidates
between data and MC, as shown in Figures \ref{fig:comparisonpipm},
\ref{fig:comparison} and \ref{fig:mks}, respectively.
Here, $U_{\rm miss}$ is required to be within $(-0.06, +0.06)$ GeV,
which includes about 98\% of
$D^{+}\rightarrow\bar K^0(\pi^{+}\pi^{-})\mu^{+}\nu_{\mu}$ and
86\% of
$D^{+}\rightarrow\bar K^0(\pi^{0}\pi^{0})\mu^{+}\nu_{\mu}$ signals.
In these figures, we can see good agreement between data and MC.

\section{Systematic uncertainty}

The common systematic uncertainty in
${\mathcal B}(D^+\to \bar K^0\mu^+\nu_\mu)$ measured with
$\bar K^0\rightarrow\pi^{+}\pi^{-}$ and $\bar K^0\rightarrow\pi^{0}\pi^{0}$
arises from
the uncertainties in the fits to the $M_{\rm BC}$ distributions,
the $\Delta E$ and $M_{\rm BC}$ requirements,
the $\mu^+$ tracking,
the $\mu^+$ PID,
the $E^{\rm extra~\gamma}_{\rm max}$ requirement,
the $M_{\bar K^0\mu^{+}}$ requirement and
the $U_{\rm miss}$ fit.
The uncertainty in the fits to the $M_{\rm BC}$ distributions is estimated to be 0.5\%
by examining the relative change of the yields of data and MC
via varying the fit range, the combinatorial background shape or the endpoint of the ARGUS function.
To estimate the uncertainties in the $\Delta E$ and $M_{\rm BC}$ requirements,
we examine the branching fractions by
enlarging the $\Delta E$ windows by 5 or 10 MeV
and
varying the $M_{\rm BC}$ windows by $\pm1$ MeV$/c^2$, respectively.
The maximum changes of the branching fractions,
which are 0.3\% and 0.3\% for $\Delta E$ and $M_{\rm BC}$ requirements,
are assigned as the uncertainties, respectively.
The uncertainties in the
tracking and PID for $\mu^+$ are estimated by analyzing $e^+e^-\to\gamma\mu^+\mu^-$ events.
The differences of the two-dimensional (momentum and $\cos\theta$) weighted
tracking efficiencies of data and MC are determined to be
$(+0.2\pm0.5)\%$ and $(-1.5\pm0.5)\%$, respectively.
We assign 0.5\% and 0.5\% as the systematic uncertainties in the tracking and PID for $\mu^+$
after correcting for these differences, respectively.
Due to different topologies, there may be difference between the weighted
efficiencies for the muons in $D^+\to \bar K^0\mu^+\nu_\mu$ and $e^+e^-\to \gamma\mu^+\mu^-$.
This difference, which is estimated to be 0.5\% by analyzing the two kinds of signal MC events,
is considered as a systematic uncertainty.
By examining the doubly tagged hadronic $D\bar{D}$ decays,
we find that the difference of the acceptance efficiencies with $E_{\rm max}^{\rm extra~\gamma}<0.15$ GeV
of data and MC is $(+3.6\pm0.1)\%$.
So, we assign 0.1\% as the uncertainty in the $E_{\rm max}^{\rm extra~\gamma}$ requirement
after correcting the MC efficiency to data.
The uncertainty in the $M_{\bar K^0\mu^{+}}$ requirement is estimated to be 0.8\%
by comparing the branching fractions measured
with alternative requirements of $M_{\bar K^0\mu^{+}}<$ 1.55 and 1.65 GeV$/c^{2}$
with the nominal value.
The uncertainty in the $U_{\rm miss}$ fit is estimated to be 0.8\%
by comparing the branching fractions measured using different
signal shape, background shape and fit range with the nominal value.
Here, to examine the uncertainty in the background shape,
we vary the relative strengths of each of the components in the inclusive MC sample
and shift the estimated numbers of other peaking backgrounds by $1\sigma$.
In our previous work, the uncertainty in the signal MC generator is
estimated to be 0.1\%, which is obtained by comparing the DT efficiencies
before and after re-weighting the $q^2 = (p_D-p_K)^2$ distribution of the signal
MC events of $D^0\to K^-e^+\nu_e$ to data~\cite{bes3-kev},
where the $p_D$ and $p_K$ are the momenta of $D$ and $K$ mesons.
Adding these in quadrature, we obtain the total common systematic uncertainty
$\delta_{\rm sys}^{\rm com}$ to be 1.6\%.

For the measurement with $\bar K^0\rightarrow\pi^{+}\pi^{-}$,
the independent systematic uncertainty arises from the uncertainties in the $\bar K^0 \rightarrow\pi^{+}\pi^{-}$
reconstruction, the MC statistics (0.4\%),
and ${\mathcal B}(\bar K^0\to \pi^+\pi^-)$ (0.1\%)~\cite{pdg2014}.
The uncertainty in the $\bar K^0 \rightarrow\pi^{+}\pi^{-}$ reconstruction is estimated
to be 1.5\% by studying
$J/\psi\to K^{*\mp}K^{\pm}$ and $J/\psi\to \phi \bar K^0 K^{\pm}\pi^{\mp}+c.c.$ events \cite{bes3-k0ev}.
Adding these uncertainties in quadrature, we obtain the total independent systematic uncertainty
($\delta_{\rm sys}^{\rm ind}$) for
$\bar K^0\rightarrow\pi^{+}\pi^{-}$ mode to be 1.6\%.

For the measurement with $\bar K^0\rightarrow\pi^{0}\pi^{0}$,
the independent systematic uncertainty arises from the uncertainties in the $\pi^{0}$ selection,
the $\bar K^0$ mass window,
the MC statistics (0.5\%),
${\mathcal B}(\bar K^0\to \pi^0\pi^0)$ (0.2\%)~\cite{pdg2014} and
the $\chi^2_1+\chi^2_2$ selection method.
The $\pi^0$ reconstruction efficiency is verified by analyzing the hadronic decays
$D^0\to K^-\pi^+$ and $K^-\pi^+\pi^+\pi^-$ versus $\bar{D^0}\to K^-\pi^+\pi^0$
and $K_{S}^{0}(\pi^+\pi^-)\pi^0$.
The difference of the $\pi^0$ reconstruction efficiencies of data and MC
is found to be $(-1.1\pm1.0)\%$ per $\pi^0$.
After correcting the detection efficiency of the signal side for this difference,
the systematic uncertainty in $\pi^0$ reconstruction is taken as
$1.0\%$ per $\pi^0$.
Here, the photons from the $\bar K^0\to K^0_S(\pi^0\pi^0)$ decays
are reconstructed under an assumption that the $K^0_S$ meson decayed at the IP.
We investigate the DT efficiencies of two kinds of signal MC events,
in which the lifetimes of $K^0_S$ meson from the signal side are set at the nominal value and 0,
respectively.
Their difference is less than 0.2\%, which is considered as the systematic uncertainty of
the $K^0_S(\pi^0\pi^0)$ reconstruction.
To avoid the effect of the $D^+ \to \bar K^0\pi^+\pi^0$ peaking backgrounds,
the uncertainty in the $\bar K^0(\pi^0\pi^0)$ mass window is estimated by examining
the ${\mathcal B}(D^+ \to \bar K^0e^+\nu_e)$ using
the same $\bar K^0(\pi^0\pi^0)$ selection criteria.
We compare the branching fractions measured using alternative $\bar K^0(\pi^0\pi^0)$
mass windows $(0.460, 0.505)$, $(0.470, 0.500)$, $(0.480, 0.500)$ GeV/$c^2$ with the
nominal value. The maximum change of the re-measured
branching fractions 0.9\% is taken as the systematic uncertainty.
The uncertainty in the $\chi^2_1+\chi^2_2$ selection method is estimated to be 0.3\%,
which is the difference of the $\pi^0\pi^0$ acceptance efficiencies of
the hadronic decays of $D^0\to K^-\pi^+\pi^0$ versus $\bar D^0\to K^+\pi^-\pi^0$
between data and MC.
Adding these in quadrature, we obtain the total independent systematic uncertainty
($\delta_{\rm sys}^{\rm com}$) for $\bar K^0\rightarrow\pi^{0}\pi^{0}$ mode to be 2.3\%.

\begin{table}[htp]
\centering \caption{Systematic uncertainties (\%) in the measurement of
${\mathcal B}(D^{+}\rightarrow\bar K^{0} \mu^{+}\nu_{\mu})$.
$\delta_{\rm sys}^{\rm com}$ and
$\delta_{\rm sys}^{\rm ind}$ denote the common and independent systematic uncertainties
for $+-$ and $00$ modes.
}
\label{tab:sys}
\begin{tabular}{ccc}
  \hline
  {\footnotesize Common source} & \multicolumn{2}{c}{\footnotesize Uncertainty} \\
  \hline
  {\footnotesize $M_{\rm BC}$ fit} & \multicolumn{2}{c}{\footnotesize 0.5} \\
  {\footnotesize $\Delta E$ requirement} & \multicolumn{2}{c}{\footnotesize 0.3} \\
  {\footnotesize ${\rm M}_{\rm BC}\in(1.863,1.877)$ GeV/$c^2$} & \multicolumn{2}{c}{\footnotesize 0.3} \\
  {\footnotesize $\mu^{+}$ tracking} & \multicolumn{2}{c}{\footnotesize 0.5} \\
  {\footnotesize $\mu^{+}$ PID} & \multicolumn{2}{c}{\footnotesize 0.5} \\
  {\footnotesize Topology difference} & \multicolumn{2}{c}{\footnotesize 0.5} \\
  {\footnotesize $E^{\rm extra~\gamma}_{\rm max}<0.15$ GeV} & \multicolumn{2}{c}{\footnotesize 0.1} \\
  {\footnotesize $M_{\bar K^0\mu^{+}}<1.6$ GeV/$c^2$} & \multicolumn{2}{c}{\footnotesize 0.8} \\
  {\footnotesize $U_{\rm miss}$ fit} & \multicolumn{2}{c} {\footnotesize 0.8} \\
  {\footnotesize MC generator} & \multicolumn{2}{c} {\footnotesize 0.1} \\
  \hline
  {\footnotesize $\delta_{\rm sys}^{\rm com}$}  & \multicolumn{2}{c}{\footnotesize 1.6} \\ \hline
  \hline
  {\footnotesize Independent source}  & {\footnotesize $\bar K^{0}\rightarrow\pi^{+}\pi^{-}$} & {\footnotesize $\bar K^{0}\rightarrow\pi^{0}\pi^{0}$} \\
  \hline
  {\footnotesize $\pi^{0}$ reconstruction}& {\footnotesize ---} & {\footnotesize 2.0}  \\
  {\footnotesize $\bar K^0(\pi^+\pi^-)$ reconstruction}  &  {\footnotesize 1.5} & ---  \\
  {\footnotesize $\bar K^0(\pi^0\pi^0)$ reconstruction}  &  {\footnotesize ---} & 0.2 \\
  {\footnotesize $M_{\pi^0\pi^0}\in(0.45, 0.51)$ GeV/$c^{2}$}  &  --- & {\footnotesize 0.9}  \\
  {\footnotesize MC statistics} & {\footnotesize 0.4}  &  {\footnotesize 0.5} \\
  {\footnotesize Quoted ${\mathcal B}(\bar K^0\to\pi\pi)$} & {\footnotesize 0.1} & {\footnotesize 0.2}   \\
  {\footnotesize $\chi^2_1+\chi^2_2$ selection method} & {\footnotesize ---} & {\footnotesize 0.3}   \\
  \hline
  {\footnotesize $\delta_{\rm sys}^{\rm ind}$}  & {\footnotesize 1.6}  &  {\footnotesize 2.3} \\
  \hline
\end{tabular}
\end{table}

Table \ref{tab:sys} summarizes the systematic uncertainties in the measurement of
${\mathcal B}(D^{+}\rightarrow\bar K^{0} \mu^{+}\nu_{\mu})$.
Quadratically combining the independent uncertainties for $+-$ and $00$ modes after
considering their observed DT yields as weights, we obtain the independent uncertainty
to be 1.4\%. Adding the common and independent uncertainties in quadrature yields the total
systematic uncertainty 2.1\%.

\section{Branching fraction}

The branching fraction of $D^{+}\rightarrow\bar K^0\mu^{+}\nu_{\mu}$
is determined by
\begin{equation}\label{equ:branchingfraction}
{\mathcal B}(D^+\to \bar K^0\mu^+\nu_\mu) =
\frac{N^{\rm prd}_{\rm DT}}
{N^{\rm tot}_{\rm ST}},
\end{equation}
where $N^{\rm prd}_{\rm DT}$
is the DT production yield corrected for detection efficiency and daughter decay branching fractions, which
has been constrained to be the same for $+-$ and $00$ modes in the simultaneous fits, and
$N^{\rm tot}_{\rm ST}$ is the total ST yield.

Inserting the numbers of $N^{\rm prd}_{\rm DT}$ and
$N^{\rm tot}_{\rm ST}$
in Eq. (\ref{equ:branchingfraction}), we obtain
\begin{equation}\label{equ:Bresult2}
{\mathcal B}(D^{+}\rightarrow\bar K^0\mu^{+}\nu_{\mu}) = (8.72 \pm 0.07 \pm 0.18)\%,
\end{equation}
where the first uncertainty is statistical and the second
systematic.

\begin{figure}[htp]
\centering
\includegraphics[width=2.8in]{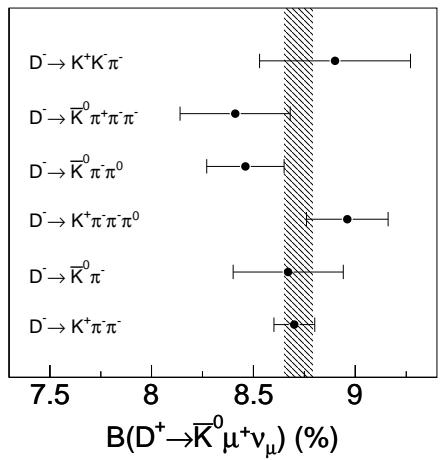}
\caption{ Comparison of the branching fractions. Dots with error
bars are results measured using different ST modes, and shadow band
is the nominal result. Only statistical uncertainties are shown.}
\label{fig:bfcompare}
\end{figure}

Furthermore, we examine the measured branching
fractions for $D^{+}\rightarrow\bar K^0\mu^{+}\nu_{\mu}$ by
separately using each of the ST modes, which are shown Fig.
\ref{fig:bfcompare}. We can see that they are consistent with
the nominal result within uncertainties very well.
Here, the uncertainties are statistical only.
The average branching fraction over the six ST modes,
weighted by their statistical uncertainties, is $(8.70\pm0.07)\%$
and is consistent with our nominal result.

\section{Summary and discussion}

In summary, by analyzing 2.93 fb$^{-1}$ of data collected at $\sqrt{s}=$ 3.773
GeV with the BESIII detector, we measure
the absolute branching fraction ${\mathcal B}(D^{+}\rightarrow\bar K^0\mu^{+}\nu_{\mu})
= (8.72 \pm 0.07_{\rm stat.} \pm 0.18_{\rm sys.})\%$,
which is consistent with previous measurements within uncertainties but with significantly improved precision.
Combining the ${\mathcal B}(D^{+}\rightarrow\bar K^0\mu^{+}\nu_{\mu})$
measured in this work with the
$\tau_{D^0}$, $\tau_{D^+}$, ${\mathcal B}(D^0\to K^-\mu^+\nu_\mu)$
and ${\mathcal B}(D^{+}\rightarrow\bar K^0e^{+}\nu_{e})$ taken from the world average \cite{pdg2014},
we determine the ratios of the partial widths
$\Gamma(D^0\to K^-\mu^+\nu_\mu)/\Gamma(D^{+}\rightarrow\bar K^0\mu^{+}\nu_{\mu})=0.963\pm0.044,$
which supports isospin conservation holding in the exclusive semi-muonic decays of $D^{+}$ and $D^{0}$ mesons,
and
$\Gamma(D^{+}\rightarrow\bar K^0\mu^{+}\nu_{\mu})/\Gamma(D^{+}\rightarrow\bar K^0e^{+}\nu_{e})=0.988\pm0.033,$
which is consistent with the predicted value in Ref.~\cite{zphyc} within uncertainties.

\section{Acknowledgements}
The BESIII collaboration thanks the staff of BEPCII and the IHEP computing center for their strong support.
This work is supported in part by National Key Basic Research Program of China under Contract Nos.
2009CB825204 and 2015CB856700;
National Natural Science Foundation of China (NSFC) under Contracts Nos. 10935007,
11125525, 11235011, 11305180, 11322544, 11335008, 11425524; the Chinese Academy of Sciences
(CAS) Large-Scale Scientific Facility Program; the CAS Center for Excellence in
Particle Physics (CCEPP); the Collaborative Innovation Center for Particles and Interactions
(CICPI); Joint Large-Scale Scientific Facility Funds of the NSFC and CAS under Contracts Nos.
11179007, U1232201, U1332201; CAS under Contracts Nos. KJCX2-YW-N29, KJCX2-YW-N45;
100 Talents Program of CAS; National 1000 Talents Program of China;
INPAC and Shanghai Key Laboratory for Particle Physics and Cosmology;
German Research Foundation DFG under Contract No. Collaborative Research
Center CRC-1044; Istituto Nazionale di Fisica Nucleare, Italy;
Koninklijke Nederlandse Akademie van Wetenschappen (KNAW) under Contract No. 530-4CDP03;
Ministry of Development of Turkey under Contract No. DPT2006K-120470;
National Natural Science Foundation of China (NSFC) under Contracts Nos.
11405046, U1332103; Russian Foundation for Basic Research under Contract No.
14-07-91152; The Swedish Resarch Council; U. S. Department of Energy under
Contracts Nos. DE-FG02-04ER41291, DE-FG02-05ER41374, DE-SC0012069, DESC0010118;
U.S. National Science Foundation; University of Groningen (RuG) and the
Helmholtzzentrum fuer Schwerionenforschung GmbH (GSI), Darmstadt;
WCU Program of National Research Foundation of Korea under Contract No. R32-2008-000-10155-0.

\end{document}